\documentclass{PoS}

\usepackage{epsfig,cite}
\usepackage{amsmath,amssymb}

\newcommand{\rmO}{\mathrm{O}}
\PoS{PoS(LAT2005)232}

\title{Restoring chiral symmetry to $\rmO(a^2)$ for dynamical Wilson fermions
\footnote{Preprint: HU-EP-05/48, SFB/CPP-05-50, DESY 05-171}}

\ShortTitle{Restoring chiral symmetry to $\rmO(a^2)$ for dynamical Wilson fermions }

\author{
\vspace{-0.5cm}
\vbox{
\epsfxsize=2.8 true cm
\epsfbox{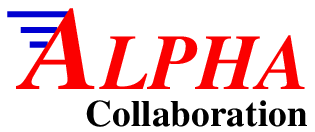}}
\vspace{0.3cm}
\speaker{Roland Hoffmann}\\
		University of Colorado,
		Boulder, CO 80309\\
		E-mail: \email{hoffmann@pizero.colorado.edu}}
\author{Michele Della Morte,
		Francesco Knechtli and Ulli Wolff\\
        HU Berlin, Newtonstr. 15, 12489 Berlin, Germany\\
        E-mail: \email{\{dellamor,knechtli,uwolff\}@physik.hu-berlin.de}}
\author{Rainer Sommer\\
		DESY, Platanenallee 6, 15738 Zeuthen, Germany\\
        E-mail: \email{sommer@ifh.de}}

\abstract{We present results for the non--perturbative determination
of the improvement and renormalization factors
of the isovector axial current for lattice QCD with two flavors of dynamical Wilson quarks.
The improvement and normalization conditions are formulated in terms of matrix elements
of the PCAC relation in the Schr\"odinger functional setup and results are given in the form
of interpolating formulae for bare gauge couplings $\beta=6/g_0^2>5.2$.\\[1cm]
Work supported by the DFG
in the Graduiertenkolleg
GK271 and the SFB/TR 09-03.}

\FullConference{XXIIIrd International Symposium on Lattice Field Theory\\
		 25-30 July 2005\\
		 Trinity College, Dublin, Ireland}

\newcommand{\be}{\vspace*{-1mm}\begin{equation}}
\newcommand{\ee}{\end{equation}\vspace*{-1mm}}
\newcommand{\bea}{\vspace*{-1mm}\begin{eqnarray}}
\newcommand{\eea}{\end{eqnarray}\vspace*{-1mm}}
\newcommand{\fa}{f_{\rm A}}
\newcommand{\fp}{f_{\rm P}}

\newcommand{\vecx}{{\bf x}}
\newcommand{\vecy}{{\bf y}}

\newcommand{\op}{\mathcal{O}}
\newcommand{\zetabar}{\bar\zeta}
\newcommand{\rmd}{{\rm d}}

\newcommand{\eq}[1]{eq.~(\ref{#1})}
\newcommand{\fig}[1]{Fig.~\ref{#1}}
\newcommand{\tab}[1]{Table~\ref{#1}}

\newcommand{\cA}{c_{\rm A}}
\newcommand{\csw}{c_{\rm SW}}
\newcommand{\za}{Z_{\rm A}}
\newcommand{\half}{{\textstyle \frac12}}

\begin{document}

\section{Introduction}

With the Wilson fermion formulation chiral symmetry is explicitly broken
at finite lattice spacing and the consequences of this explicit
breaking have to be dealt with. Among those is the fact that $\rmO(a)$
counter terms in the Symanzik expansion are no longer excluded and that there
is no conserved Noether current associated with a continuous chiral symmetry
of the lattice action.

The first problem is addressed by the Symanzik improvement programme, i.e. the
addition of irrelevant operators to both the action and the composite fields.
By tuning the coefficients of only a few terms it is possible to remove
the $\rmO(a)$ scaling violations from the theory. Here we compute the improvement
coefficient for the isovector axial current $\cA(g_0^2)$, which -- together
with the improvement of the action \cite{Jansen:1998mx} -- will ensure the absence
of $\rmO(a)$ lattice artifacts in \emph{all} matrix elements of the PCAC relation,
which involve insertions at finite separation
\cite{Luscher:1996sc}.

The normalization factor $\za(g_0^2)$ of the improved axial current is obtained
by enforcing a continuum--like transformation behavior under infinitesimal chiral rotations.
Since the isovector chiral symmetry is recovered in the continuum and the normalization
condition is based on a local identity, $\za$ is finite and scale--independent.

For unexplained notation and additional details about the improvement and normalization
conditions we refer to \cite{DellaMorte:2005se} and \cite{DellaMorte:2005rd}, respectively.

\section{Axial current improvement}

When calculating a bare quark mass on the lattice from matrix elements of the
PCAC relation
\be
\partial_\mu A^a_\mu(x)=2mP^a(x)\;,\label{pcac}
\ee
any dependence on the\vspace*{0.4mm} kinematic parameters or external states, i.e. the dependence
on the precise
choice of matrix\vspace*{-0.4mm} element, is a cutoff effect.
A non--perturbative improvement condition can thus be obtained by inserting into
\eq{pcac} the improved axial current \cite{Luscher:1996sc}
\be
(A_{\rm I})_\mu^a(x)=A_\mu^a(x)+a\cA\half(\partial_\mu+\partial^*_\mu)P^a(x)\;,
\label{impr}
\ee
and tuning $\cA$ such that\vspace*{0.4mm} the masses obtained from two different matrix elements
agree. In (\ref{impr}) $\partial_\mu$ ($\partial_\mu^*$) denotes the forward (backward)
lattice\vspace*{-0.4mm} derivative.

When evaluating improvement coefficients non--perturbatively
one has to keep in mind that due to cutoff effects in the correlation
function used to formulate the improvement condition, the coefficients themselves
are uncertain to $\rmO(a)$.\footnote{For the same reason $\za$
is uncertain to $\rmO(a^2)$ after improvement.}
While this forbids a unique definition of the improved theory,
the $\rmO(a)$ ambiguities can be made to disappear smoothly
if the improvement condition is evaluated with all physical scales kept
fixed, 
while only the lattice spacing is varied \cite{Guagnelli:2000jw}.
In the evaluation of our improvement condition
we keep the bare quark mass (using the 1--loop value for
$\cA$ from \cite{Luscher:1996vw}) constant, thus ignoring small changes
in renormalization factors, and fix the relative lattice spacing
in the range of bare gauge couplings we consider through asymptotic
scaling \cite{DellaMorte:2005se}.

It is also important to make sure that the correlation functions in the
improvement condition are not dominated by states with energy close to
the cutoff. If this were the case, the improvement condition might
cancel exceptionally large scaling violations and a $\cA$
obtained in this way could introduce significant $\rmO(a^2)$ effects.

We construct matrix elements of (\ref{pcac}) between pseudo--scalar
states and the vacuum in the Schr\"odinger functional
\cite{Luscher:1992an,Sint:1993un}. More precisely, in this work we use
\bea 
 \fa(x_0;\omega)&=&- \frac{a^3}{3L^6} \sum_\vecx\langle A_0^a(x) \, \op^a(\omega)
\rangle\;,\\[-2mm]
\textrm{and }\ \     \fp(x_0;\omega)&=&- \frac{a^3}{3L^6} \sum_\vecx\langle   P^a(x) \, \op^a(\omega)
\rangle\;,\\[-6mm]\nonumber
\eea
with the pseudo--scalar operator
\be
        \quad \op^a(\omega) =a^6 
         \sum_{\vecx,\vecy} \zetabar(\vecx) \gamma_5 \tau^a\half 
        \omega(\vecx-\vecy) \zeta(\vecy)\;.\label{op}
\ee
It lives at the $x_0\!=\!0$ boundary of the\vspace*{0.4mm} SF cylinder and depends on a spatial 
trial ``wave function'' $\omega$. A quark mass
from (\ref{pcac}) using (\ref{impr}) is then given by $m=r+a\cA s+\rmO(a^2)$
with
\bea
r(x_0;\omega)=\frac{\half(\partial_0+\partial^*_0)\fa(x_0;\omega)}
{2\fp(x_0;\omega)}\quad
&\textrm{and}&\quad s(x_0;\omega)=\frac{\partial_0\partial^*_0 \fp(x_0;\omega)}
{2\fp(x_0;\omega)}\;.\label{sSF}
\eea
In our chosen setup\vspace*{0.4mm} we now have $x_0$, the insertion time, and $\omega$, the spatial
trial wave function, as parameters for probing the PCAC relation. Enforcing the\vspace*{-0.5mm}
independence of the quark mass from these results in
\be
-\cA=\frac{\Delta r}{a\Delta s}=\frac1a\!\cdot\!\frac{r(x_0;\omega_{\pi,1})
-r(y_0;\omega_{\pi,0})}
{s(x_0;\omega_{\pi,1})-s(y_0;\omega_{\pi,0})}\;,\label{cadef}
\ee
and therefore the sensitivity to $\cA$ is given by $a\Delta s$.\vspace*{0.4mm}

The combined requirement of large sensitivity to $\cA$ and explicit
control over excited--state contributions is met with the method
proposed in \cite{Durr:2003nc} where it is tested in the quenched
approximation. There $\omega_{\pi,0/1}$ from (\ref{cadef})
are chosen such that $\op(\omega_{\pi,0/1})$ excite mostly the
ground and first excited state in the pseudo--scalar channel, respectively.
Thus the energy of the states can be monitored directly using the effective
mass of $f_{\rm A/P}$ and the sensitivity
$a\Delta s=as(x_0,\omega_{\pi,1})-as(x_0,\omega_{\pi,0})\propto {m_\pi^\star}^2-m_\pi^2$
is also expected to be large.

The simulations listed in \tab{t_simpar}
are at constant physical volume according to asymptotic scaling as described in
\cite{DellaMorte:2005se} and the bare (unimproved) PCAC mass is kept constant to
$\simeq10\%$.
\STABULAR[htb]{|llll|}{\hline
$L/a$  &  $\beta$ & $\ m\cdot L$ & $-\cA$\\\hline
12   & 5.20 &  $0.18(1)$  & 0.0638(23)\\
16   & 5.42 &  $0.200(5)$ & 0.0420(21)\\
24   & 5.70 &  $0.182(5)$ & 0.0243(36)\\\hline}
{Simulation parameters for $\cA$.\label{t_simpar}}
We simulate a set of three spatial wave functions and approximate the combinations
$\op(\omega_{\pi,0/1})$ that project to the ground and first excited states
through the eigenvectors of the correlation matrix
\cite{Durr:2003nc,DellaMorte:2005se}
\bea 
f_1(\omega',\omega)&=&- \frac1{3L^6}           \langle   \op'^a (\omega') \,
\op^a(\omega) \rangle\;,
\eea
where $\op'$ is a pseudo--scalar operator at the $x_0=T$ boundary. In \fig{effmass}
\vspace*{0.6mm}
two distinct signals are clearly visible, which indicates that the
approximate projection method works well at our parameters.
The energy of the first excited state is not far away from $a^{-1}$,
suggesting that in even smaller volumes the residual $\rmO(a^2)$ effects would
grow rapidly.

\DOUBLEFIGURE[t]{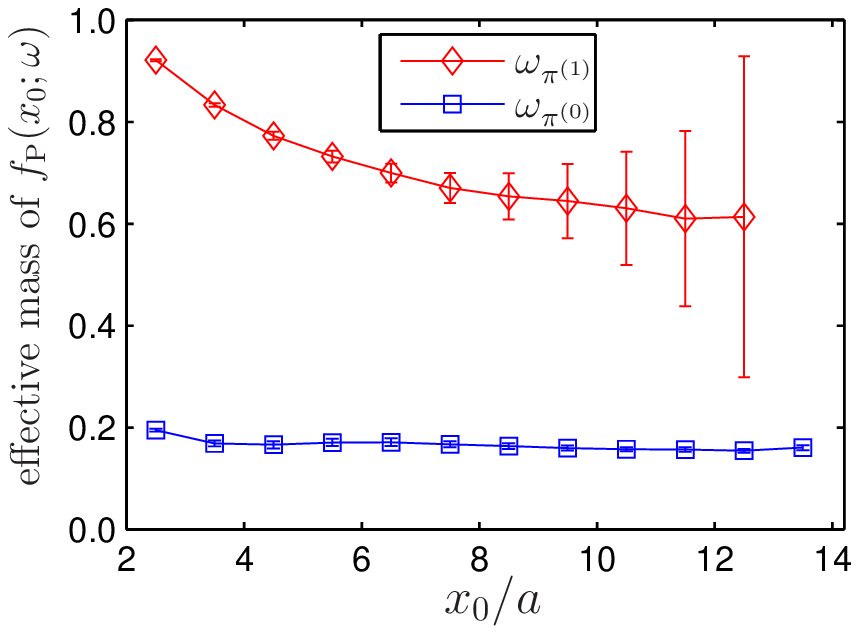,height=4.5cm}{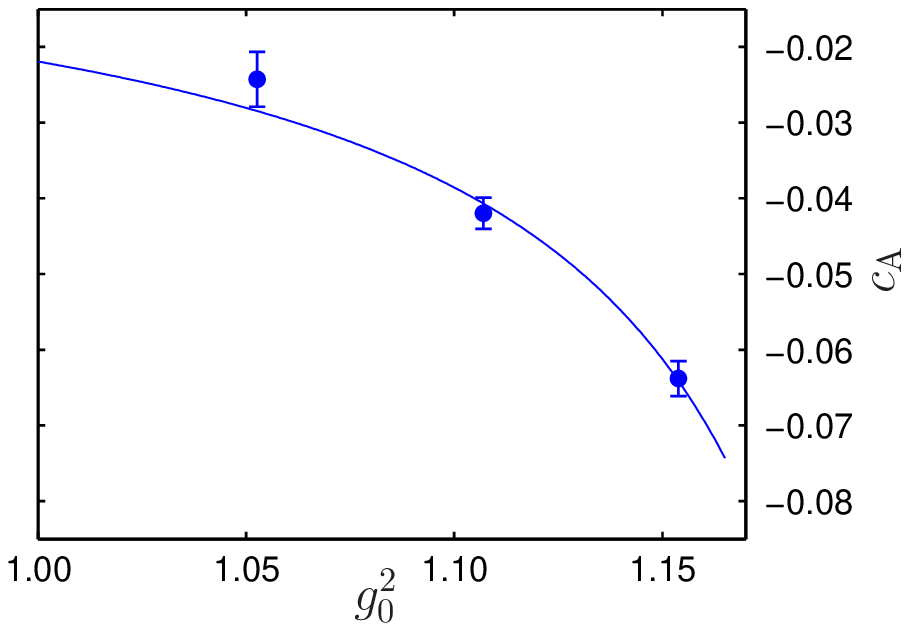,height=4.32cm}{
Effective pseudo--scalar masses of $\fp$ from the $\beta=5.42$ run.\label{effmass}}
{Simulation results for $\cA$. The solid line is given by the interpolating
formula (\protect\ref{caint}).
\label{final}}

Our definition of $\cA$ is completed by fixing $x_0=y_0=T/2$ in (\ref{cadef}) and specifying
$L=T$ and $\theta=0$. The results
from the simulations summarized in \tab{t_simpar} are shown in \fig{final} as a function of $g_0^2$.
The solid line is a smooth interpolation of the simulation data,
constrained in addition by 1--loop perturbation
theory:\\[-5mm]
\begin{equation}
\cA(g_0^2)=-0.00756\, g_0^2\times\frac{1-0.4485\, g_0^2}{1-0.8098\, g_0^2}\;.
\label{caint}
\end{equation}
It is our final result, valid in the range $0.98 \leq g_0^2 \leq1.16$  
within the errors of the data points (at most 0.004).
By performing additional simulations we have verified that the volumes in our runs
are scaled sufficiently precisely such that systematic errors due to deviations from
the constant physics condition can be neglected. The same also
applies to variations in the quark mass.\\[-5mm]

\section{Axial current renormalization}

In a massless renormalization scheme preservation of $\rmO(a)$--improvement implies
that the renormalized improved current is of the form \cite{Jansen:1995ck,Luscher:1996sc}
\be
(A_{\rm R})_\mu^a=\za(1+b_{\rm A}m_{\rm q})(A_{\rm I})_\mu^a\;.\label{aren}
\ee
The normalization condition we use \cite{Hoffmann:2003mm,DellaMorte:2005rd} is based
\vspace*{0.8mm}
on \cite{Luscher:1996jn}, the ALPHA collaboration's quenched determination of $\za$.
Since a massless scheme requires the normalization condition to be set up at vanishing
quark mass, in \cite{Luscher:1996jn} the mass term of the axial Ward identity was dropped
in the derivation of the normalization condition. In practice this condition shows a very
pronounced quark mass dependence and thus a potential extrapolation is rather steep and
the location of the critical point must be known with high precision.

Performing a chiral transformation in the continuum and keeping track of the mass term
results in the integrated Ward identity \cite{DellaMorte:2005rd}
\begin{eqnarray}
\int\!\!\rmd^3\vecy\,\rmd^3\vecx\, \epsilon^{abc}
\Big\langle A_0^a(y_0\!+\!t,\vecx)A_0^b(y)\op_{\rm ext}\Big\rangle
\qquad\qquad\nonumber\\
-2m\!\int\!\!\rmd^3\vecy\,\rmd^3\vecx\!\int_{y_0}^{y_0+t}\!\!\!\!\!\rmd x_0\,
\epsilon^{abc}\Big\langle P^a(x)A_0^b(y)\op_{\rm ext}\Big\rangle
&\!=\!&i\displaystyle\!\int\!\!\rmd^3\vecy\
\Big\langle V_0^c(y)\op_{\rm ext}\Big\rangle\;\qquad\label{WI3}
\end{eqnarray}
and the choice\footnote{No use of spatial wave functions is made in the axial
current renormalization, i.e. here $\op\equiv\op(\omega=const)$.}
$\op^c_{\rm ext}=-\epsilon^{cde}\op'^d\op^e/{6L^6}$
allows us to replace the r.h.s. of (\ref{WI3}) with $f_1$ by using isospin
symmetry. A normalization condition is then obtained by requiring
that (\ref{WI3}) holds for the renormalized improved current (\ref{aren}).

\fig{fig:extrapol} shows the results of an evaluation of this condition
at a bare gauge coupling of $\beta=5.2$. To show the effect of the inclusion
of the mass term, also the results with the method from \cite{Luscher:1996jn}
are shown, i.e. when dropping the second term on the l.h.s. of (\ref{WI3}).
While for the new normalization condition
the slope in $am$ is consistent
with zero\footnote{Any remnant slope is due to the neglect of $b_{\rm A}$ as
well as contact terms in the second term in the l.h.s. of (\ref{WI3}).},
the estimate of $\za$ from the old condition changes
by $30\%$ in the (small) mass range shown.
We anyway see that for $am\lesssim0.02$ all
mass effects show a linear behavior.
For $\beta=5.5$ the extrapolation is similar to
the one shown and at all other gauge couplings
we can in fact interpolate from two simulations
very close to the critical point.
\STABULAR[b]{|c|rrll|}{
    \hline
$\beta$ &
\multicolumn{1}{c}{$L/a$} &
\multicolumn{1}{c}{$T/a$} &
\multicolumn{1}{c}{$\kappa_c\qquad$} &
\multicolumn{1}{c|}{$\za\qquad$} \\
    \hline
5.200 &  8 & 18& 0.135856(18)& 0.7141(123)     \\
5.500 & 12 & 27& 0.136733(8) & 0.7882(35)(39)  \\
5.715 & 16 & 36& 0.136688(11)& 0.8037(38)(7)   \\\hline
5.290 &  8 & 18& 0.136310(22)& 0.7532(79)      \\\hline
7.200 &  8 & 18& 0.134220(21)& 0.8702(16)(7)   \\
8.400 &  8 & 18& 0.132584(7) & 0.8991(25)(7)   \\
9.600 &  8 & 18& 0.131405(3) & 0.9132(11)(7)   \\\hline
}
{Results for the chiral extrapolations of $\za$ and estimates for the critical hopping
parameter $\kappa_c$.\label{z_simpar}}
In addition to
three (again asymptotically) matched lattice sizes at
$L/a=8,12$ and $16$, we simulated at
three larger values of $\beta$ and fixed $L/a=8$, which
corresponds to very small volumes.
This was done in
order to verify that our non--perturbative estimate
smoothly connects to the perturbative predictions
\cite{Gabrielli:1990us,
Gockeler:1996gu}. The first error we quote for $\za$
is statistical and the second represents our estimate
of the systematic error, which originates from deviations from
the constant physics condition.
\DOUBLEFIGURE[t]{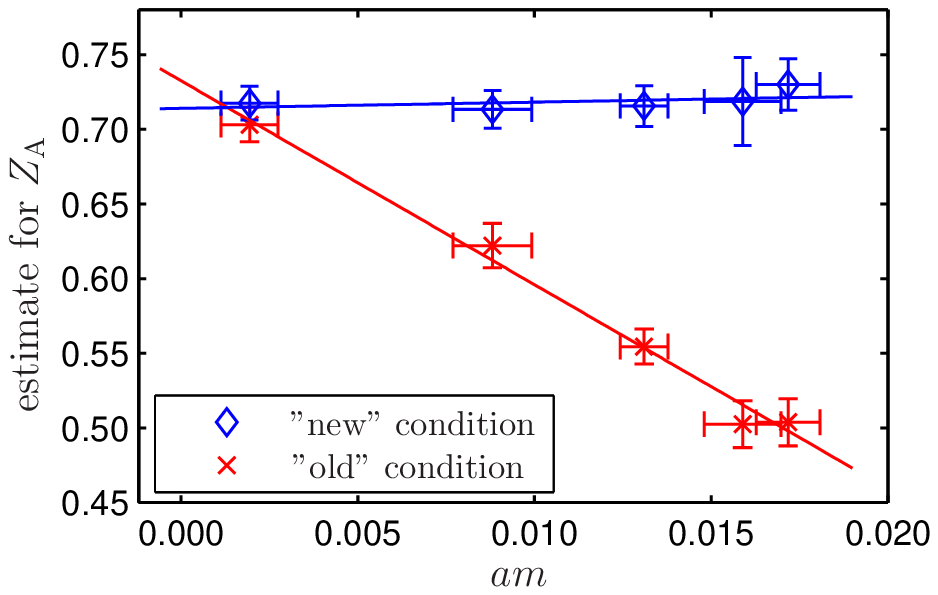,height=4.5cm}{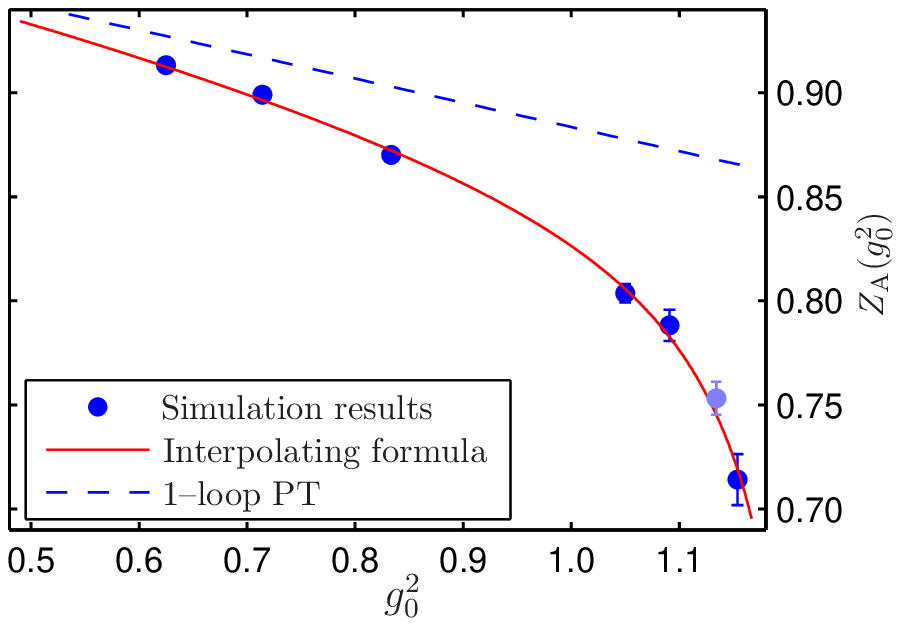,height=4.22cm}{
Comparison of the chiral extrapolation at $\beta=5.2$
using the new and old normalization conditions for $\za$.\label{fig:extrapol}}
{Result for $\za$ from numerical
simulations and 1--loop perturbation theory
(dashed line). The Pade fit (solid lines)
is given by (\protect\ref{zaint}).\label{fig:result}}
There is no estimate of the systematic error for the $\beta=5.29$ run, which was done
only to verify the rapid change of $\za$ in this region of bare gauge coupling.
It is thus also excluded from a fit, which results in the interpolating formula
(again incorporating the 1--loop asymptotic constraint)
\be
\za(g_0^2)=\frac{1 -0.918\,g_0^2+0.062\,g_0^4+0.020\,g_0^6}
{1 -0.8015\,g_0^2}\;.\label{zaint}\\[-6mm]
\ee

\section{Summary and Outlook}

For the $\rmO(a)$-improved action with non-perturbative $\csw$
\cite{Jansen:1998mx},
we have determined the improvement coefficient $\cA$ 
for $\beta\!\geq\!5.2$, which roughly corresponds to $a\!\leq\!0.1$~fm. 
The improvement condition was implemented at constant physics, which is 
necessary in a situation when $\rmO(a)$ ambiguities in the improvement
coefficients are not negligible.

Through the calculation of $\za(g_0^2)$ we have shown that in a lattice
theory with two flavors of Wilson fermions normalization
conditions can be imposed at the non--perturbative level
such that isovector chiral symmetries are realized
in the continuum limit.
Since we are working with an improved theory, chiral
Ward--Takahashi identities are then satisfied up to
$\rmO(a^2)$ at finite lattice spacing.

The improvement and normalization conditions were implemented in terms
of correlation functions in the Schr\"odinger functional
framework and evaluated
on a line of constant physics in order to achieve
a smooth disappearance of the $\rmO(a)$ and $\rmO(a^2)$ uncertainties.
Clearly, the methods employed in this paper may also be useful to
compute $\cA$ and $\za$ in the three flavor case, where $\csw$ is known
non--perturbatively with plaquette and Iwasaki gauge actions 
\cite{Aoki:2002vh,Yamada:2004ja,Aoki:2005et}. 

The determinations of $\cA$ and $\za$ were carried
out within the ALPHA collaboration's programme to calculate quark
masses in a fully non--perturbative framework. The results of
this programme and its application to the strange quark mass
are presented in \cite{DellaMorte:2005kg,michele_proceedings}.

{\renewcommand{\baselinestretch}{0.86}
\bibliography{../refs}
\bibliographystyle{JHEP-2}}

\end{document}